\documentclass[prc,twocolumn,superscriptaddress,preprintnumbers,showpacs,showkeys,floatfix]{revtex4}
\usepackage{graphicx}

\def\bfK{{\bf K}}

\def\be{\begin{equation}}
\def\ee{\end{equation}}
\def\bea{\begin{eqnarray}}
\def\non{\nonumber\\}
\def\eea{\end{eqnarray}}

\newcommand{\eq}[1]{Eq.~(\ref{#1})}

\newcommand{\bmath}[1]{\mbox{\boldmath${#1}$}}
\newcommand{\brho}{\bmath{\rho}}

\newcommand{\ki}[1]{\bmath{k}_{#1}^{}}
\newcommand{\kij}[2]{\bmath{k}_{#1}^{#2}}
\newcommand{\ri}[1]{\bmath{r}_{#1}^{}}
\newcommand{\rij}[2]{\bmath{r}_{#1}^{#2}}
\newcommand{\qi}[1]{\bmath{q}_{#1}^{}}
\newcommand{\ppi}[1]{\bmath{p}_{#1}^{}}

\def\bfR{\bmath{R}}
\def\bfS{\bmath{S}}
\def\bfr{\bmath{r}}
\def\bfK{\bmath{K}}
\def\bfk{\bmath{k}}
\def\bfl{\bmath{l}}
\def\bfp{\bmath{p}}
\def\bfv{\bmath{v}}

\newcommand{\bkappa}{\bmath{\kappa}}

\newcommand{\bsigma}{\bmath{\sigma}}\newcommand{\bSigma}{\bmath{\Sigma}}

\newcommand{\expup}[1]{e^{#1}}

\date{\today}
\begin{document}
\vskip1in\vskip1in\vskip1in\vskip1in
\preprint{NT@UW-07-03}

\title{Initial-State Coulomb Interaction in the
\boldmath$dd\to\alpha\pi^0_{}$ Reaction}

\author{Timo A. L\"ahde}
\email{talahde@u.washington.edu}
\affiliation{Department of Physics, University of Washington, Seattle, 
WA 98195-1560}
\author{Gerald A. Miller}
\email{miller@phys.washington.edu}
\affiliation{Department of Physics, University of Washington, Seattle, 
WA 98195-1560}

\begin{abstract}
The effects of initial-state Coulomb interactions in the charge-symmetry-breaking 
reaction \mbox{$dd\to\alpha\pi^0_{}$} are investigated within a previously published formalism. 
This is a leading order effect in which the Coulomb interaction between the two 
initial state protons leads to the breakup of the two deuterons into a continuum 
state that is well connected to the final $\alpha\pi^0_{}$ state by the strong
emission of a pion. As a first step, we use a simplified set of $d$ and 
$\alpha$ wave functions and a plane-wave approximation for the initial $dd$ state. 
This Coulomb mechanism, by itself, yields cross sections that are much larger than the 
experimental ones, and which are comparable in size to the contributions from other 
mechanisms. Inclusion of this mechanism is therefore necessary in a realistic
calculation.
\end{abstract}


\maketitle

\section{Introduction}
The concepts of charge independence and charge symmetry provide powerful tools in
organizing the multiplet structure of systems of hadrons and nuclei.
These symmetries are not perfect; diverse small but interesting violations have
been discovered; see the reviews of Refs.~\cite{MNS,MOS}. Our concern here is with 
charge symmetry breaking.

\begin{figure*}[t]
\includegraphics[width=.8\textwidth]{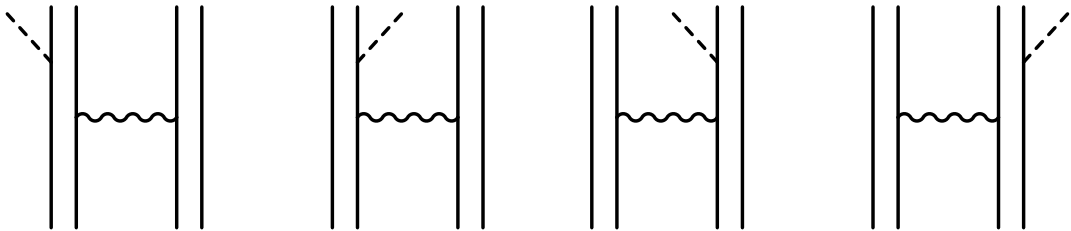}
\caption{Diagrams relevant for the inclusion of the Coulomb initial-state 
interaction. The wiggly lines represent the Coulomb interaction between
the two protons of the initial state. The dashed lines represent the emitted pion.
For each Coulomb interaction, anyone of the four nucleons may emit the pion.}
\label{diag}
\end{figure*}

Hadronic states can be regarded as approximately charge symmetric, i.e.~invariant 
under a rotation by $180^\circ$ around the 2-axis in isospin space. Charge 
symmetry~(CS) is a subset of the general isospin symmetry, charge 
independence~(CI), which requires invariance under \emph{any} rotation in isospin 
space. In Quantum Chromodynamics~(QCD), CS requires that the dynamics are unchanged 
under the exchange of the up and down quarks~\cite{MNS}. In the language of 
hadrons, this symmetry translates into e.g.~the invariance of the strong interaction 
under the exchange of protons and neutrons. However, since the up and down quarks 
do have different masses~($m_u^{}\neq m_d^{}$)~\cite{wei77,mumd}, the QCD Lagrangian is 
not charge symmetric and neither are the strong interactions of hadrons. 
This symmetry violation is called charge symmetry breaking~(CSB). The different
electromagnetic interactions of the up and down quarks also contribute to CSB.

Observing the effects of CSB interactions therefore provides a probe of $m_u^{}$ and 
$m_d^{}$, which are fundamental, but poorly known, parameters of the standard model. 
For example, the light quark mass difference causes the neutron to be heavier
than the proton. If this were not the case, our universe would be very different, 
as a consequence of the dependence of Big-Bang nucleosynthesis on the relative 
abundances of protons and neutrons. Experimental evidence for CSB has been 
demonstrated many times, see e.g. Refs.~\cite{MNS,MOS}. Two exciting recent 
observations of CSB in experiments involving the production of neutral pions have 
stimulated current interest: Many years of effort led to the observation of CSB in 
$np\to d\pi^0$ at TRIUMF. After a careful treatment of systematic errors, the CSB 
forward-backward asymmetry of the differential cross section was found to be 
$A_{\rm fb}^{}=(17.2 \pm 8 \pm 5.5)\times 10^{-4}_{}$~\cite{Allena}, where the 
former error is statistical and the latter systematical. In addition, the final 
experiment at the IUCF Cooler ring has reported a very convincing $dd\to\alpha\pi^0$ 
signal near threshold ($\sigma=12.7\pm2.2$~pb at $T_d^{}=228.5$~MeV and $15.1 \pm 3.1$~pb 
at $231.8$~MeV)~\cite{IUCFCSB}. These data are consistent with the pion being produced 
in an $s$-wave, as expected from the proximity of the threshold ($T_d^{}=225.6$~MeV).
Studies of the $dd\to\alpha\pi^0$ reaction thus present exciting new opportunities 
for developing the understanding of CSB. 

The reaction $dd\to\alpha\pi^0$ obviously violates isospin conservation, but more 
specifically, it violates charge symmetry since the deuterons and the $\alpha$-particle 
are self-conjugate under the charge-symmetry operator, with a positive eigenvalue, 
while the neutral pion wave function changes sign. This reaction could thus not occur 
if charge symmetry were conserved, and its cross section is proportional to the square 
of the CSB amplitude. This phenomenon is unique, because all other observations of CSB 
involve interferences with charge-symmetric amplitudes.

Due to the recent availability of high-quality experimental data on CSB, a theoretical 
interpretation using fundamental CSB mechanisms is called for. At momenta comparable 
to the pion mass, $Q\sim m_\pi^{}$, QCD and its symmetries (and in particular CSB) can 
be described by a hadronic effective field theory~(EFT), called Chiral Perturbation 
Theory~($\chi$PT), for extensive reviews see Refs.~\cite{ulfreview,birareview,hanhartreview}.
This EFT has been extended to pion production in Refs.~\cite{cpt0,cpt3,Rocha,cpt1,cpt2}
where typical momenta are $Q\sim \sqrt{m_\pi^{}M}$, with $M$ as the nucleon 
mass (see also Ref.~\cite{pionprod} where pion production was studied neglecting 
this large momentum in power counting). The EFT formalism provides specific CSB effects 
in addition to the nucleon mass difference. In particular, there are two pion-nucleon 
seagull interactions related by chiral symmetry to the quark-mass and electromagnetic 
contributions to the nucleon mass difference~\cite{vkiv,wiv}. 

In previous work~\cite{first,realistic}, the cross section for the reaction $dd\to\alpha\pi^0$
was computed near threshold by chiral EFT techniques, using a chiral power counting scheme
to assess the expected importance of different interaction terms. The first paper~\cite{first} 
used a plane wave 
approximation along with Gaussian bound-state wave functions. These initial calculations 
yielded computed cross sections that are a factor of $\sim 2$ larger than the measured ones.
The effects of initial-state interactions and realistic bound-state wave functions
were included later~\cite{realistic}, with resulting cross sections of the order of  
several hundred pb or more. It is thus clear that more work is needed to understand the 
order of magnitude of the measured cross section, such as a treatment of the effects of 
loop diagrams along the lines of Ref.~\cite{Lensky}.

We consider here one specific contribution that has previously been 
neglected -- the influence of the Coulomb interaction in the initial state. The formalism 
employed is similar to that of Ref.~\cite{first}. The main objective is to assess the 
relevance of this mechanism, so we use simple bound-state wave functions and neglect the 
effects of strong interactions in the initial state. It is worthwhile to explain some basic 
features of the calculation that result from invariance principles: Spin, isospin, and 
symmetry requirements restrict the partial waves allowed for the $dd\to\alpha\pi^0$ 
reaction. In the spectroscopic notation $^{2S+1}\!L_J^{}l$, where $S, L$ and $J$ are,
respectively, the spin, orbital, and total angular momenta of the $dd$ state, and $l$ 
denotes the pion angular momentum, the lowest allowed partial waves are $^3\!P_0^{}s$ 
and $^5\!D_1^{}p$. Hence, production of an $s$-wave pion requires that the initial 
deuterons be in a relative $P$-wave, with spins coupled to a spin-1 state, forming
together a state with zero total angular momentum. 
On the other hand, a $p$-wave pion is produced only when the deuterons are in a
relative $D$-wave, with spins maximally aligned to spin 2, requiring either a 
coupling with $\Delta L=\Delta S=2$ or $D$-states of $d$ or $\alpha$.
Interferences between $s$- and $p$-waves disappear for any unpolarized 
observable. We shall therefore be concerned with the production of an $s$-wave pion.
In the mechanism proposed here, the Coulomb interaction between the two protons of
different deuterons converts the initial relative $P$-wave state into an $S$-wave state.
Parity conservation then requires that one of the resulting pairs of nucleons be in
a $p$-wave with orbital angular momentum unity. The strong pseudovector pion production 
operator then converts this $p$-wave state into an $s$-wave state.

This paper is organized in the following manner: The relevant Coulomb mechanism is 
described in Sect.~\ref{mech}, which also explains how the formalism of 
Ref.~\cite{first} is to be employed. We use the simple Gaussian bound-state wave 
functions of Ref.~\cite{first}, but extend the calculation by also considering
Hulth\'en wave functions for the deuteron. The detailed evaluation and numerical 
results are given in Sect.~\ref{eval}. For comparison, the effects of 
Coulomb interactions in the final state are considered in Sec.~\ref{sec:final}. 
These have been calculated in Ref.~\cite{realistic} and were found to be 
very small. Finally, Sect.~\ref{incorp} assesses our results and discusses how these 
effects can be included in a realistic calculation that incorporates the strong 
interactions in the initial state.

\section{Coulomb Mechanism}
\label{mech}

In the present study, CSB arises from the initial-state Coulomb interaction between
the two deuterons, followed by strong pion emission, as shown in Fig.~\ref{diag}. 
We note that the electromagnetic contributions can be 
ordered~\cite{first} relative to each other in the same fashion as the effects of 
strong CSB. In this case, the leading order~(LO) term considered here is of 
$\mathcal{O}\left[\alpha_{\mathrm{em}}^{} M/(4\pi f_\pi^3 p)\right]$, with $M$ 
as the nucleon mass. 
This term is roughly of the same size as the LO strong CSB term which is of 
$\mathcal{O}\left[{m_d^{}-m_u^{}\over m_d^{}+m_u^{}} 
\,m_\pi^2/(M 4\pi f_\pi^3 p)\right]$. 

The CSB pion production operator ${\cal O}_{C}^{}$ is given by
\begin{eqnarray} 
{\mathcal O}_{C}^{} &=& {\mathcal O}_1^{}
\left(E-H_0^{}+i\epsilon\right)^{-1}_{} V_C^{}, 
\label{op}
\end{eqnarray} 
where $V_C^{}$ is the Coulomb interaction between the two protons in the initial
state, which acts to form a four-body continuum state that propagates
according to \mbox{$(E-H_0^{}+i\epsilon)^{-1}$.} It is convenient to write 
$V_C^{}$ as a sum of pair-wise operators:
\bea 
V_C^{} &=& \sum_{j<k=1,4} 
Q_j^{} Q_k^{} v_C^{j,k},
\eea
where the $Q_{j,k}^{}$ are nucleon charge operators. The operator 
$H_0^{}$ is the sum of the kinetic energies of each of the four nucleons. 
The strong pion production operator is denoted by ${\cal O}_1^{}$ and is given by
\begin{eqnarray}
\mathcal{O}_{1}^{} &=& \frac{g_A^{}}{2f_\pi^{}}
\sum_{i}\tau_{z,i}^{}\:\bsigma_i^{}\!\cdot\! 
\left[\qi{i}-\frac{\omega}{2M}(\kij{i}{'}+\ki{i})\right] \nonumber \\
&\rightarrow& \left(-\frac{g_A^{}}{2f_\pi^{}}\right) \frac{\mu}{M}
\sum_{i}\tau_{z,i}^{}\:\bsigma_i^{}\!\cdot \ki{i},
\end{eqnarray}
where the $\ki{i},\kij{i}{'}$ are nucleon momenta before and after
the pion emission, respectively. The $p$-wave term with $\qi{i}=-\ppi{\pi}$ 
can be ignored in the threshold regime considered here. The factor $\omega$ 
is also replaced by the pion mass $\mu = 134.974$~MeV.

The present analysis uses a plane-wave approximation and the simplest 
possible $d$ and $\alpha$ bound-state wave functions, those of a Gaussian form. 
Assuming spatially symmetric bound-state wave functions, the
invariant amplitude is given by
\begin{eqnarray}
\mathcal{M} & = & \int d^3r d^3\!\rho_1 d^3\!\rho_2\:
\langle A|\mathcal{O}|DD\rangle,
\label{amp}
\end{eqnarray}
with
\begin{eqnarray}
|A\rangle & = & \sqrt{2E_\alpha^{}}\:
\Psi_\alpha^{}(\bfr,\brho_1^{},\brho_2^{})\:|\alpha\rangle, \\
|DD\rangle & = & \sqrt{s}\:\Phi_d^{}(\brho_1^{})\Phi_d^{}(\brho_2^{})\:|dd\rangle,
\end{eqnarray}
where $\Psi_{\alpha}^{}$ and $\Phi_d^{}$ are the spatial parts of the 
$\alpha$-particle and deuteron bound-state wave functions, and $s=4E_d^2$ is 
the total c.m. energy squared. The ket vectors $|\alpha\rangle$ and $|dd\rangle$ 
contain the fully anti-symmetrized spin and isospin wave functions.
Because of symmetry requirements, the plane-wave $dd$ scattering wave
function is included in $|dd\rangle$ as given by Eqs.~(\ref{eq:dd}) 
and~(\ref{eq:ddexp}) below. The invariant amplitude can then be written as
\begin{eqnarray}
\mathcal{M} & = & \sqrt{2E_\alpha^{}s}\int d^3r d^3\!\rho_1^{} d^3\!\rho_2^{}
\:\Psi^{\dagger}_{\alpha}(\bfr,\brho_1^{},\brho_2^{}) \nonumber \\ 
&& \langle\alpha|\mathcal{O}|dd\rangle\:\Phi_d^{}(\brho_1^{})\Phi_d(\brho_2^{}),
\end{eqnarray}
The matrix element $\langle\alpha|\mathcal{O}|dd\rangle$ contains all the spin-isospin 
couplings of the nucleons and the pion production operator $\mathcal{O}$. The 
wave functions are expressed in terms of the (2+2) Jacobian coordinates
\begin{eqnarray}
\bfR & = & \frac{1}{4}(\ri{1}+\ri{2}+\ri{3}+\ri{4}) \quad 
(\equiv 0\ {\rm in\ c.m.}), \nonumber \\
\bfr & = & \frac{1}{2}(\ri{1}+\ri{2}-\ri{3}-\ri{4}), \nonumber \\
\brho_1^{} & = & \ri{1}-\ri{2}, \nonumber \\
\brho_2^{} & = & \ri{3}-\ri{4},
\end{eqnarray}
with the corresponding momenta
\begin{eqnarray}
\bfK & = & \ki{1}+\ki{2}+\ki{3}+\ki{4} \quad 
(\equiv 0\ {\rm in\ c.m.}), \nonumber \\
\bfk & = & \frac{1}{2}(\ki{1}+\ki{2}-\ki{3}-\ki{4}) \nonumber \\
& = & \frac{1}{2}(\ppi{1}-\ppi{2}) \quad 
(\equiv \bfp\ {\rm in\ c.m.}), \nonumber \\
\bkappa_1^{} & = & \frac{1}{2}(\ki{1}-\ki{2}), \nonumber \\
\bkappa_2^{} & = & \frac{1}{2}(\ki{3}-\ki{4}), 
\end{eqnarray}
defined so that $\sum_i \ki{i}\cdot\ri{i}=\bfK\cdot\bfR+\ki{}\cdot\bfr+
\bkappa_1^{}\cdot\brho_1^{}+\bkappa_2^{}\cdot\brho_2^{}$. The Jacobians 
are equal to unity in both representations.

The ground-state wave functions are represented by Gaussian functions, and
these may be explicitly expressed in the above coordinates using 
$\sum_{i<j}(\ri{i}-\ri{j})^2=4\rij{}{2}+2\brho_1^2+2\brho_2^2$, yielding
\begin{eqnarray}
\Psi_{\alpha}^{}(\bfr,\brho_1^{},\brho_2^{}) & \!=\! & 
\frac{8}{\pi^{9/4}\alpha^{9/2}}\:\exp\left[
-\frac{\left(2\bfr^2\!+\!\brho_1^2\!+\!\brho_2^2\right)}{\alpha^2_{}}\right], 
%\quad \:\: \\
\quad\quad \\
\Phi_{d}^{}(\brho_i^{}) & \!=\! & \frac{1}{\pi^{3/4}\beta^{3/2}}\:
\exp\left(-\frac{\brho_i^2}{2\beta^2_{}}\right), \:\: i=1,2 \quad
\end{eqnarray}
where the parameters \mbox{$\alpha=2.77$~fm} and \mbox{$\beta=3.189$~fm} 
have been fixed using the measured $\alpha$ and $d$ rms point radii 
$\langle r_\alpha^2\rangle^{1/2}=1.47$~fm and 
$\langle r_d^2\rangle^{1/2}=1.953$~fm~\cite{radii}. We shall work in momentum 
space and therefore record the corresponding wave functions
\begin{eqnarray}
\widetilde{\Psi}_\alpha(\bfk,\bkappa_1^{},\bkappa_2^{}) &=& N_\alpha^{}\: 
\exp{\left[-\frac{\alpha^2}{8}(\bfk^2+2\bkappa_1^2+2\bkappa_2^2)\right]},
\nonumber \\
N_\alpha^{} &\equiv& {\alpha^{9/2}\over 8\pi^{9/4}} \label{alpha}, \\
\widetilde{\Phi}_d^{}(\kappa_i^{}) &=& N_d^{}\:
\exp{\left(-{\bkappa_i^2\beta^2_{} \over 2}\right)}, \quad i=1,2 \quad \nonumber \\
N_d &\equiv& \left({\beta^2\over\pi}\right)^{3/4}.
\label{deut}
\end{eqnarray}
In order to study the sensitivity of our results to the choice of wave functions,
we also use a deuteron wave function of the Hulth\'en form:
\begin{eqnarray}
\widetilde{\Phi}_{d}^H(\bkappa_i^{}) &=& N_d^H
\left(\frac{1}{\bkappa_i^2+a^2_{}}-\frac{1}{\bkappa_i^2+b^2_{}}\right), 
\quad i=1,2 \quad \nonumber \\
N_d^H &=& \frac{\sqrt{ab(a+b)}}{\pi(a-b)},
\label{hulth}
\end{eqnarray} 
where the parameters are given by 
\mbox{$a=0.23161$~fm$^{-1}$} and \mbox{$b=1.3802$~fm$^{-1}$}~\cite{Tiburzi}.

Since we have assumed that the orbital parts of the wave functions are 
symmetric under the exchange of any pair of nucleons, we may define the 
initial- and final-state spin-isospin wave functions as
\begin{eqnarray}
\left|\alpha\right\rangle & = & \frac{1}{\sqrt{2}}\left\{ 
\left((1,2)_1^{},(3,4)_1^{}\right)_0^{}
\left[[1,2]_0^{},[3,4]_0^{}\right]_0^{} 
\right. \nonumber \\
&& \left. -
\left((1,2)_0^{},(3,4)_0^{}\right)_0^{}
\left[[1,2]_1^{},[3,4]_1^{}\right]_0^{} 
\right\}, \nonumber \\
&\equiv& \frac{1}{\sqrt{2}}
\left(\left|\alpha_1^{}\right\rangle + \left|\alpha_2^{}\right\rangle\right) 
\label{eq:alpha} \\
\left|dd\right\rangle & = & \frac{1}{\sqrt{3}}\left(
1-P_{23}^{}-P_{24}^{}\right)\,\left|d_{12}^{}d_{34}^{}\right\rangle,
\label{eq:dd} \\
\left|d_{12}^{}d_{34}^{}\right\rangle & = & 
\left((1,2)_1^{},(3,4)_1^{}\right)_S^{}
\left[[1,2]_0^{},[3,4]_0^{}\right]_0^{} \nonumber \\
&& \times \frac{1}{\sqrt{2}}\left[\expup{i\bfp\cdot\bfr} +
(-)^L_{}\expup{-i\bfp\cdot\bfr}\right],
%&& \times \frac{1}{\sqrt{2}}\left[\expup{i\bp\cdot\ri{}} +
%(-)^L_{}\expup{-i\bp\cdot\ri{}}\right],
\label{eq:ddexp}
\end{eqnarray}
where $(i,j)_s^{}$ and $[i,j]_T^{}$ are the spin and isospin Clebsch-Gordan 
couplings, with magnetic quantum numbers suppressed, for nucleons, or nucleon 
pairs, $i$ and $j$ coupling to spin $s$ and isospin $T$, respectively. 
We shall refer to the first term of Eq.~(\ref{eq:alpha}) as the ``$dd$'' component 
of the $\alpha$ because the pairs~(12) and~(34) each have the spin and isospin 
of the deuteron. In the above equations, $P_{ij}^{}$ is the permutation 
operator of the indicated nucleons. The symmetry requirements for the exchange 
of the deuterons are represented by the (orbital angular momentum dependent) 
combination of plane waves in Eq.~(\ref{eq:ddexp}), with $\bfp$ as the relative 
momentum of the deuterons. Even though the expression for the $\alpha$ state 
superficially singles out a (12)+(34) configuration, it is indeed 
fully anti-symmetric in all indices. This particular form is used because it 
closely matches the form of the initial-state wave functions, thereby simplifying 
the evaluation of the spin-isospin summations in the matrix element. In practice,
the $dd$ wave function can be simplified to
\begin{eqnarray}
\left|dd\right\rangle & = & \sqrt{6}\,
\left((1,2)_1^{},(3,4)_1^{}\right)_S^{} 
\left[[1,2]_0^{},[3,4]_0^{}\right]_0^{}\:
e^{i\bfp\cdot\bfr}, \quad\quad
\label{ddi}
\end{eqnarray}
since each of the three terms in Eq.~(\ref{eq:dd}) gives an identical 
contribution to the matrix element.

The expressions~(\ref{eq:alpha}) through~(\ref{ddi}) provide insight that 
simplifies the calculation: The Coulomb interaction has no spin operator, so the 
initial state is connected only with the ``$dd$'' component of the alpha particle,
which means that only the Class~III~\cite{MNS} part of the Coulomb operator 
$(\tau_{z,i}^{} +\tau_{z,j}^{})$ contributes. One of these $\tau_z^{}$ operators
finds another from within the pion production operator and is squared to $\pm$~unity.
As a result, the matrix element turns out to be proportional to the spin operators 
of the~(1,2) and~(3,4) systems. In the normalization used here, the spin-averaged 
cross section (for $s$-wave pions) is given by
\begin{eqnarray}
\sigma &=& \frac{1}{16\pi s}\frac{|\ppi{\pi}|}{|\bfp|}\:
\frac{1}{9}\sum_{\rm pol.}|\mathcal{M}|^2_{},
\label{sigma}
\end{eqnarray}
where the summation is over the deuteron polarizations.

\section{Evaluation}
\label{eval}

The analysis is most conveniently performed in momentum space. Combination 
of~\eq{op} with~\eq{amp} yields, upon Fourier transformation,
the Coulomb contribution of present interest, ${\cal M}_C^{}:$
\bea 
{\cal M}_C^{} &=& \sqrt{E_\alpha^{} s}
\left(-\frac{g_A^{}}{2f_\pi^{}}\right){\mu\over M}
\int d^3k\,d^3\kappa_1^{}d^3\kappa_2^{} \nonumber \\
&\times& \langle\alpha_1^{} 
\vert\widetilde{\Psi}_\alpha(\bfk,\bkappa_1^{},\bkappa_2^{})\sum_{i=1,4} 
\frac{\tau_{z,i}^{}\:\bsigma_i^{}\cdot \ki{i}}
{E-{2\bkappa_1^2+2\bkappa_2^2+\bfk^2\over 2M}+i\epsilon} \nonumber\\
&\times& (\bfk,\bkappa_1^{},\bkappa_2^{}\vert V_C^{}
\vert dd, \bfp\rangle,
\label{matel1}
\eea
where the relation $\sum_{j=1,4}{\bfk_j^2} = {2\bkappa_1^2+2\bkappa_2^2+\bfk^2}$
has been applied in the free propagator, and $\vert dd,\bfp\rangle$ represents the 
initial $dd$ (relative plane wave) state of Eq.~(\ref{ddi}), including the internal 
spatial, spin, and isospin degrees of freedom. Further, 
$(\bfk,\bkappa_1^{},\bkappa_2^{}\vert V_C^{}\vert dd,\bfp\rangle$ denotes the momentum 
space representation of the state formed by the action of $V_C^{}$ on the initial 
state. The round bracket notation used here signifies that only the spatial 
degrees of freedom are included.

\subsection{Reduction to Quadrature}
The first step in the calculation is the simplification of the pion
production operator in Eq.~(\ref{matel1}). Define the operator 
$X$ according to
%${g_A^{}\mu\over f_\pi^{} M}X\;(=G_0^{-1}{\cal O}_C^{})$:
\begin{widetext}
\bea
X &=& \sum_{i,j<k=1,4} 
\tau_{z,i}^{}\:\bsigma_i^{}\cdot \ki{i} \:Q_j^{}Q_k^{}\: v_C^{j,k} \\
&=& (\tau_{z,3}^{}\:\bsigma_3^{}\cdot\bfk_3^{}
+\tau_{z,4}^{}\:\bsigma_4^{}\cdot\bfk_4^{})
\:\frac{(1+\tau_{z,3}^{})}{2}\:
\left[\frac{(1+\tau_{z,1}^{})}{2}\:v_C^{3,1}
+\frac{(1+\tau_{z,2}^{})}{2}\:v_C^{3,2}\right]\non
&+& (\tau_{z,3}^{}\:\bsigma_3^{}\cdot\bfk_3^{}
+\tau_{z,4}^{}\:\bsigma_4^{}\cdot\bfk_4^{})
\:\frac{(1+\tau_{z,4}^{})}{2}\:
\left[\frac{(1+\tau_{z,1}^{})}{2}\:v_C^{4,1}
+\frac{(1+\tau_{z,2}^{})}{2}\:v_C^{4,2}\right]\non
&+& (\tau_{z,1}^{}\:\bsigma_1^{}\cdot\bfk_1^{}
+\tau_{z,2}^{}\:\bsigma_2^{}\cdot\bfk_2^{})
\:\frac{(1+\tau_{z,1}^{})}{2}\:
\left[\frac{(1+\tau_{z,3}^{})}{2}\:v_C^{3,1}
+\frac{(1+\tau_{z,4}^{})}{2}\:v_C^{4,1}\right]\non
&+& (\tau_{z,1}^{}\:\bsigma_1^{}\cdot\bfk_1^{}
+\tau_{z,2}^{}\:\bsigma_2^{}\cdot\bfk_2^{})
\:\frac{(1+\tau_{z,2}^{})}{2}\:
\left[\frac{(1+\tau_{z,3}^{})}{2}\:v_C^{3,2}
+\frac{(1+\tau_{z,4}^{})}{2}\:v_C^{4,2}\right]
\label{one}.
\eea 
It is instructive to consider the first term of~\eq{one}: It should be noted that 
the operators $v_C^{3,1},v_C^{3,2}$ do not flip the spin of their deuteron. Also, 
the initial-state deuteron~$(1,2)$ is connected to the deuteron-like~$(1,2)$ 
component of the $\alpha$. Thus the terms with $\tau_{z,1}^{},\tau_{z,2}^{}$ can
be dropped, and the initial-state deuteron~$(3,4)$ is similarly connected to the 
deuteron-like~$(3,4)$ component of the $\alpha$. We are required to have CSB, so
only the terms proportional to $\tau_{z,3}^{}$ are relevant. For the~$(3,4)$ 
``deuteron'' of the final state we have $\tau_{z,3}^{}=-\tau_{z,4}^{}$. Thus the 
first term of~\eq{one} simplifies to $\frac{1}{4} 
(\bsigma_3^{}\cdot\bfk_3^{}-\bsigma_4^{}\cdot\bfk_4^{})(v_C^{3,1}+v_C^{3,2})$.
Similar manipulation of the remaining terms in~\eq{one} leads to the result

\bea
X &=& \left(\frac{\bsigma_3^{}\cdot\bfk_3^{}-\bsigma_4^{}\cdot\bfk_4^{}}{4}\right)
\left[v_C^{3,1}+v_C^{3,2}-v_C^{4,1}-v_C^{4,2}\right]
\:+\: \left(\frac{\bsigma_1^{}\cdot\bfk_1^{}-\bsigma_2^{}\cdot\bfk_2^{}}{4}\right)
\left[v_C^{3,1}+v_C^{4,1}-v_C^{3,2}-v_C^{4,2}\right].
\eea
Next define the spin operators $\bfS_1^{}={1\over2}(\bsigma_1^{}+\bsigma_2^{}), 
\bfS_2^{}={1\over2}(\bsigma_3^{}+\bsigma_4^{}),
\bSigma_1^{}={1\over2}(\bsigma_1^{}-\bsigma_2^{}), 
\bSigma_2^{}={1\over2}(\bsigma_3^{}-\bsigma_4^{})$, such that each of the
$\bsigma_i^{}$ is a linear combination of the $\bfS_i^{}$ and $\bSigma_i^{}$. 
Only the terms proportional to $\bfS_i^{}$ connect the initial state to 
the ``$dd$'' component of the $\alpha$. Thus one finds
\bea 
X &\rightarrow& \frac{\bfS_2^{}\cdot\bkappa_2^{}}{2}
\left[v_C^{3,1}+v_C^{3,2}-v_C^{4,1}-v_C^{4,2}\right]
\:+\: \frac{\bfS_1^{}\cdot\bkappa_1^{}}{2}
\left[v_C^{3,1}+v_C^{4,1}-v_C^{3,2}-v_C^{4,2}\right].
\label{two}
\eea
Further, it is permissible to interchange indices~3 and~4 in the spatial wave 
functions multiplying the first term of~\eq{two}, and similarly to 
interchange~1 and~2 in those multiplying the second term. The final form of the
operator $X$ is thus
\bea 
X &\rightarrow& \bfS_2^{}\cdot\bkappa_2^{}
\left[v_C^{3,1}+v_C^{3,2}\right]
\:+\: \bfS_1^{}\cdot\bkappa_1^{}
\left[v_C^{3,1}+v_C^{4,1}\right].
\label{three}
\eea   

The next task is to compute the momentum space matrix element of the 
operator $X$. This is given by
\bea 
\left(\bfk,\bkappa_1^{},\bkappa_2^{}\right\vert X \left\vert dd,\bfp\right) &=&
\bfS_2^{}\cdot\bkappa_2^{}\:\left(\bfk,\bkappa_1^{},\bkappa_2^{}\right\vert 
(v_C^{3,1}+v_C^{3,2})\left\vert dd,\bfp\right)
\:+\: \bfS_1^{}\cdot\bkappa_1^{}\:\left(\bfk,\bkappa_1^{},\bkappa_2^{}\right\vert 
(v_C^{3,1}+v_C^{4,1})\left\vert dd,\bfp\right),
\label{four}
\eea
with the spatial matrix elements
\bea 
\left(\bfk,\bkappa_1^{},\bkappa_2^{}\right\vert v_C^{j,k}\left\vert dd,\bfp\right) &=& 
\int {d^3 \rho_1^{} d^3\rho_2^{} d^3r\over (2\pi)^{9/2}} \:\:
e^{-i\bkappa_1^{}\cdot\brho_1^{}-i\bkappa_2^{}\cdot\brho_2^{}-i\bfk\cdot\bfr}
\:\frac{\alpha_\mathrm{em}^{}}{\vert \bfr_j^{}-\bfr_k^{} \vert}\: e^{i\bfp\cdot\bfr}\:\:
\Phi_d^{}(\brho_1^{})\Phi_d^{}(\brho_2^{}),
\eea
which are thus found to be products of the momentum-space Coulomb interaction
with deuteron wave functions evaluated at shifted values of the momentum. In particular,
we define $\bfv\equiv \bfk-\bfp$ and obtain
\bea
\left(\bfk,\bkappa_1^{},\bkappa_2^{}\right\vert v_C^{3,1}\left\vert dd,\bfp\right) &=&
\frac{4\pi\alpha_{\mathrm{em}}^{}}{(2\pi)^{3/2} \bfv^2}\:\:
\widetilde{\Phi}_d^{}\left(\bkappa_1^{}-{\bfv\over2}\right)
\widetilde{\Phi}_d^{}\left(\bkappa_2^{}+{\bfv\over2}\right), \non
\left(\bfk,\bkappa_1^{},\bkappa_2^{}\right\vert v_C^{3,2}\left\vert dd,\bfp\right) &=&
\frac{4\pi\alpha_{\mathrm{em}}^{}}{(2\pi)^{3/2} \bfv^2}\:\:
\widetilde{\Phi}_d^{}\left(\bkappa_1^{}+{\bfv\over2}\right)
\widetilde{\Phi}_d^{}\left(\bkappa_2^{}+{\bfv\over2}\right), \non
\left(\bfk,\bkappa_1^{},\bkappa_2^{}\right\vert v_C^{4,1}\left\vert dd,\bfp\right) &=&
\frac{4\pi\alpha_{\mathrm{em}}^{}}{(2\pi)^{3/2} \bfv^2}\:\:
\widetilde{\Phi}_d^{}\left(\bkappa_1^{}-{\bfv\over2}\right)
\widetilde{\Phi}_d^{}\left(\bkappa_2^{}-{\bfv\over2}\right).
\label{five}
\eea
Insertion of these results into~\eq{four} finally gives
\bea
\left(\bfk,\bkappa_1^{},\bkappa_2^{}\right\vert X 
\left\vert dd,\bfp\right) &=&
\frac{8\pi\alpha_{\mathrm{em}}^{}}{(2\pi)^{3/2} \bfv^2}\:
\left(\bfS_2^{}\cdot\bkappa_2^{}-\bfS_1^{}\cdot\bkappa_1^{}\right)\:
\widetilde{\Phi}_d^{}\left(\bkappa_1^{}+{\bfv\over2}\right)
\widetilde{\Phi}_d^{}\left(\bkappa_2^{}+{\bfv\over2}\right).
\label{eight}
\eea
Examination of~\eq{matel1} reveals that $\bfp$ is the only momentum remaining after the integrals
have been performed. Thus the terms with $\bkappa_1^{}$ and $\bkappa_2^{}$ in Eq.~(\ref{eight}) both
end up being proportional to $\widehat{\bfp}$, which we may take as the $z-$axis. Furthermore, as the
integrands of both terms are identical, the whole operator must be proportional to 
$(S_{2z}^{}-S_{1z}^{})$ and the integrand to $\frac{\bkappa_1^{}+\bkappa_2^{}}{2}$. We thus need to
consider the spin matrix element
\bea 
\langle 1M_1^{},1M_2^{}\vert 0,0\rangle
\langle 1M_1^{},1M_2^{}\vert(S_{2z}^{}-S_{1z}^{})\vert 1M_1^{},1M_2^{}\rangle
&=& \langle 1M_1^{},1M_2^{}\vert 0,0\rangle (M_2^{}-M_1^{}) \nonumber \\
&=& 2M_2^{} \langle 1M_1^{},1M_2^{}\vert 0,0\rangle,
\label{nine}
\eea
where 
$\langle 1M_1^{},1M_2^{}\vert 0,0\rangle$ is the 
Clebsch-Gordan coefficient that couples the spins in the ``$dd$'' component to zero. Armed with
this knowledge, we may now use Eqs.~(\ref{ddi}) and~(\ref{eight},\ref{nine}) in the matrix 
element~(\ref{matel1}) to obtain 
\bea 
{\cal M}_C^{} &=& \sqrt{6 E_\alpha^{} s}
\left(-\frac{\mu g_A^{}}{f_\pi^{}}\right)
\frac{8\pi\alpha_{\mathrm{em}}^{}}{(2\pi)^{3/2}}
\:\times\:
M_2^{} \langle 1M_1^{},1M_2^{}\vert 0,0\rangle 
\label{matel2} \\
&\times&
\int d^3kd^3\kappa_1^{}d^3\kappa_2^{}\: 
\frac{(\bkappa_1^{}+\bkappa_2^{})\cdot\widehat{\bfp}}{\bfv^2}\:
\left[2ME-(2\bkappa_1^2+2\bkappa_2^2+\bfk^2)+i\epsilon\right]^{-1}_{}\:
\widetilde{\Psi}_\alpha(\bfk,\bkappa_1^{},\bkappa_2^{})\:
\widetilde{\Phi}_d^{}\left(\bkappa_1+\frac{\bfv}{2}\right)\:
\widetilde{\Phi}_d^{}\left(\bkappa_2+\frac{\bfv}{2}\right). \nonumber 
\eea
The above equation is our main result, and it allows for the use of general radial wave functions.
However, because of the zeros in the energy denominator, it may not be well suited to evaluation
using Monte-Carlo techniques. Nevertheless, if certain simple wave functions are used, 
Eq.~(\ref{matel2}) may be simplified further. In the next subsection, this will be performed for 
wave functions of the Gaussian and Hulth\'en types.

\subsection{Gaussian and Hulth\'en Deuteron Wave Functions}

If the Gaussian wave functions of Eqs.~(\ref{alpha}) and~(\ref{deut}) are used, the
expression~(\ref{matel2}) becomes
\bea
{\cal M}_C^{} &=& \sqrt{6 E_\alpha^{} s}\; \left(-\frac{\mu g_A^{}}{f_\pi^{}}\right)
\frac{8\pi\alpha_{\mathrm{em}}^{}}{(2\pi)^{3/2}}
\:\times\:
M_2^{} \langle 1M_1^{},1M_2^{}\vert 0,0\rangle
\:\times\:
N_a^{}N_d^2
\:\times\:
I_g^{},\label{idef} \\
I_g^{} &=& \int d^3kd^3\kappa_1^{}d^3\kappa_2^{}\:\:
\frac{(\bkappa_1^{}+\bkappa_2^{})\cdot\widehat{\bfp}}{\bfv^2}\:\: 
\left[2ME-(2\bkappa_1^2+2\bkappa_2^2+\bfk^2)+i\epsilon\right]^{-1}_{}
\nonumber \\
&\times&
\exp{\left[-\frac{\alpha^2}{8}(\bfk^2+2\bkappa_1^2+2\bkappa_2^2)\right]} \:
\exp{\left[-\frac{\beta^2}{2}\left(\bkappa_1^{}+\frac{\bfv}{2}\right)^2\right]} \:
\exp{\left[-\frac{\beta^2}{2}\left(\bkappa_2^{}+\frac{\bfv}{2}\right)^2\right]}, 
\label{matel3}
\eea
where we recall the definition $\bfv\equiv \bfk-\bfp$. The factors in the 
denominator of~\eq{matel3} may be rewritten in terms of Gaussians, giving
\bea
\bfv^{-2}_{} &=& \int_0^\infty d\gamma \:
\exp\left[-\gamma(\bfv^2 +\epsilon_1^{})\right], \\
\left[2ME-(2\bkappa_1^2+2\bkappa_2^2+\bfk^2)+i\epsilon\right]^{-1}_{} 
&=& -i\int_0^\infty d\nu\:
\exp\left[\,i\nu\left(2ME-(2\bkappa_1^2+2\bkappa_2^2+\bfk^2)+i\epsilon\right)\right],
\label{extra}
\eea
where the regulator $\epsilon$ assures that the integral over $\nu$ converges, and 
$\epsilon_1^{}$ is included to handle the point $\bfk=\bfp$. The use
of the above identities leads to an 11~dimensional integral, of which 9~dimensions involve
products of Gaussian functions, such that these integrals may be computed analytically by
successive completion of squares in the exponents. This procedure yields a two-dimensional
integral over $\nu$ and $\gamma$, which is then computed numerically. In this way, using
the definitions
\bea
&&
\bkappa \:\equiv\: \bkappa_1^{}+\bkappa_2^{}, \quad
\bfl \:\equiv\: \frac{\bkappa_1^{}-\bkappa_2^{}}{2}, \quad
\bkappa_1^{} \:=\: \frac{\bkappa}{2} + \bfl, \quad
\bkappa_2^{} \:=\: \frac{\bkappa}{2} - \bfl, \quad
\bkappa_1^2+\bkappa_2^2 \:=\: 2\bfl^2+\frac{\bkappa^2}{2}, 
\eea
the integral $I_g^{}$ may be re-written as 
\bea
I_g^{} &=& -i\int_0^\infty d\nu\,d\gamma
\int d^3v\:d^3\kappa\:d^3l
\:\:\bkappa\cdot\widehat{\bfp}\:\:
\exp{\left[-{(\bfv+\bfp)^2\alpha^2\over 8}-\left(2\bfl^2+\frac{\bkappa^2}{2}\right)
\left(\frac{\alpha^2}{4}+\frac{\beta^2}{2}\right)
\right]} \non
&\times&
\exp{\left[-\frac{\bkappa\cdot\bfv\beta^2}{2} - \frac{\bfv^2\beta^2}{4} 
- \gamma (\bfv^2+\epsilon_1^{})\right]}\:
\exp{\left[i\nu\,(2ME-4\bfl^2-\bkappa^2-(\bfv+\bfp)^2+i\epsilon)\right]},
\label{I1}
\eea
where the completion of squares is facilitated by the definitions
\bea 
&&
\bar{\alpha}^2\equiv\alpha^2+8i\nu, \quad
R_l^2 \equiv \frac{\alpha^2}{2} + \beta^2 + 4i\nu, \quad
R_v^2 \equiv \frac{R_l^2}{4} + \gamma - \frac{\beta^4}{4R_l^2}, \quad 
E = \frac{\bfp^2}{M_d^{}} = \frac{\bfp^2}{2M}.
\eea 
For all successive equations, we will define $I_j^{} = -i|\bfp|\pi^{9/2}_{}I_{1j}^{}$. At
this point, the regulators $\epsilon$ and $\epsilon_1^{}$ may be safely dropped. By 
combination of the above results and definitions, we find for the case of Gaussian
deuteron wave functions
\bea
{\cal M}_C^{} &=& \sqrt{6 E_\alpha^{} s}\:
\left(-\frac{\mu g_A^{}}{f_\pi^{}}\right)
\frac{8\pi\alpha_{\mathrm{em}}^{}}{(2\pi)^{3/2}}
\:\times\:
M_2^{} \langle 1M_1^{},1M_2^{}\vert 0,0\rangle 
\:\times\:
N_a^{}N_d^2
\:\times\:
(-i|\bfp|\pi^{9/2}_{})
\:\times\:
I_{1g}^{}, \non
I_{1g}^{} &=& \int_0^\infty d\nu\,d\gamma\:
\frac{\beta^2\bar{\alpha}^2}{R_l^8 R_v^5}\:
\exp\left\{-\bfp^2\left[\frac{\alpha^2}{8}\left(1-\frac{\alpha^2}{8R_v^2}\right)
+ \frac{\nu^2}{R_v^2}\right]\right\}\:
\exp\left(\frac{i\nu\bfp^2\alpha^2}{4R_v^2}\right).
\label{matel4}
\eea

If the Hulth\'en wave function, given in~\eq{hulth}, is used for the deuteron, the
product of deuteron wave functions in~\eq{matel2} may be re-written as
\bea
\widetilde{\Phi}_d^h\left(\bkappa_1^{}+\frac{\bfv}{2}\right) 
\widetilde{\Phi}_d^h\left(\bkappa_2^{}+\frac{\bfv}{2}\right) 
&=& \left({N_d^h}\right)^2\int_0^\infty d\eta_1^{}d\eta_2^{}
\left\{
\exp\left[-\eta_1^{}\left(\bkappa_1^{}+\frac{\bfv}{2}\right)^2\right]
\left(e^{-\eta_1^{}a^2}-e^{-\eta_1^{}b^2}\right)
\right\} \nonumber \\
&\times& \left\{
\exp\left[-\eta_2^{}\left(\bkappa_2^{}+\frac{\bfv}{2}\right)^2\right]
\left(e^{-\eta_2^{}a^2}-e^{-\eta_2^{}b^2}\right)
\right\},
\eea
after which the calculation proceeds, as before, through successive 
completion of squares, but in the Hulth\'en case we are left with a 
four-dimensional integral suitable for numerical evaluation. The end 
result is that the integral $I_{1g}^{}$ of~\eq{matel4} should be
replaced by $I_{1h}^{}$, which is given by
\bea 
I_{1h}^{} &=&
\int_0^\infty d\nu\,d\gamma\,d\eta_1^{}d\eta_2^{}\:
\frac{\eta_{12}^{}\bar{\alpha}^2_{}}{8 R_l^3 R_\kappa^5 R_v^5}\:
\exp\left\{-\bfp^2\left[\frac{\alpha^2}{8}\left(1-\frac{\alpha^2}{8R_v^2}\right)
+ \frac{\nu^2}{R_v^2}\right]\right\}\:
\exp\left(\frac{i\nu\bfp^2\alpha^2}{4R_v^2}\right)\:
f_{ab}^{}(\eta_1^{},\eta_2^{}),
\eea
where the definitions
\bea
&& \eta_{12}^{}=\frac{\eta_1^{}+\eta_2^{}}{4}-\frac{(\eta_1^{}-\eta_2^{})^2}{4R_l^2},\quad 
R_l^2=\frac{\bar{\alpha}^2}{2}+\eta_1^{}+\eta_2^{},\quad
R_\kappa^2=\frac{\bar{\alpha}^2}{8}+\eta_{12}^{},\quad
R_v^2=R_\kappa^2+\gamma-\frac{\eta_{12}^2}{R_\kappa^2}, \\
&& f_{ab}^{}(\eta_1^{},\eta_2^{}) \:=\:
e^{-a^2(\eta_1^{}+\eta_2^{})}_{} +
e^{-b^2(\eta_1^{}+\eta_2^{})}_{} - 
2\,e^{-a^2\eta_1^{}} e^{-b^2\eta_2^{}},
\eea
are used. These expressions represent the complete amplitude. In order to obtain
the cross section, it is necessary to evaluate the spin sum
\begin{eqnarray} 
\sum_{M_1^{},M_2^{}}\;M_2^2
\left\langle 1M_1^{},1M_2^{}\vert 0,0\right\rangle^2_{}
&=& \frac{2}{3}
\end{eqnarray}
and insert everything into~\eq{sigma}, yielding
\bea
\sigma_g^{} &=& E_\alpha^{}
\left(\frac{\mu g_A^{}}{f_\pi^{}}\right)^2
\frac{|\bfp||\bfp_\pi^{}|\alpha_{\mathrm{em}}^2}{288}
\:\frac{\alpha^9_{}\beta^6_{}}{\sqrt{\pi}}\:\left\vert I_{1g}^{}\right\vert^2, \\
\sigma_h^{} &=& 
E_\alpha^{}
\left(\frac{\mu g_A^{}}{f_\pi^{}}\right)^2
\frac{|\bfp||\bfp_\pi^{}|\alpha_{\mathrm{em}}^2}{288}
\:\frac{\alpha^9_{}}{\pi^{3/2}}\:
\frac{a^2b^2(a+b)^2}{(a-b)^4}\:
\left\vert I_{1h}^{}\right\vert^2,
\eea
where $\sigma_g^{}$ and $\sigma_h^{}$ again denote the expressions relevant for the Gaussian
and Hulth\'en deuteron wave functions, respectively. It should also be noted that the above 
expressions remain valid for the Coulomb interaction in the $^4_{}$He bound state, which is
considered in Sect.~\ref{sec:final}.

\section{Coulomb in the $^4_{}$He Bound State}
\label{sec:final}

\begin{figure}[t]
\includegraphics[width=.8\columnwidth]{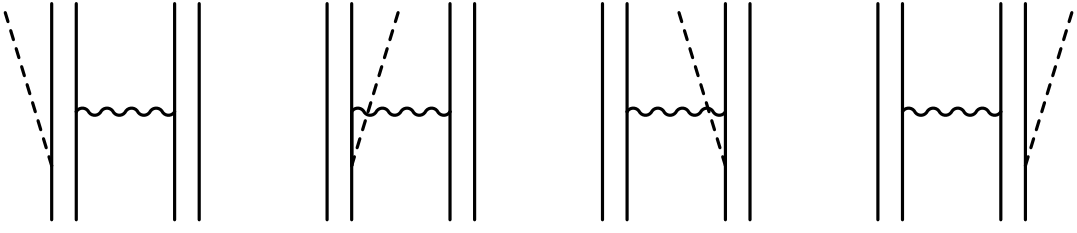}
\caption{Diagrams relevant for the inclusion of the Coulomb final-state 
interaction. The wiggly lines represent the Coulomb interaction between
the two protons in the final state. The dashed lines represent the emitted pion.}
\label{diagf}
\end{figure}

A complete assessment of all Coulomb effects should include a treatment of the Coulomb interactions
in both the initial and final states. Our focus here is on the initial-state effects, as these have
not been considered up to now. However, it is also worthwhile to compute the effects of the Coulomb
interactions in the final state within the present framework. We recall that in this framework, the
strong interaction between the initial-state deuterons is neglected, and simple bound-state 
wave functions are used. If the effects of Coulomb interactions in the $^4_{}$He bound state are 
included, the reaction $dd\to\alpha\pi^0_{}$ can proceed via strong pion production, which is here
assumed to be initiated by the one-body operator. The relevant CBS pion production operator is
then given by
\begin{eqnarray} 
{\mathcal O}_C^F &=& V_C^{} \left(-\epsilon_B^{}-H_0^{}+i\epsilon\right)^{-1}_{} 
{\mathcal O}_1^{},
\label{op1}
\end{eqnarray} 
where $\epsilon_B^{} \simeq 28.3$~MeV is the $^4_{}$He binding energy. 
It is instructive to define the operator~$Y$ according to
\bea
Y &=& \sum_{i,j<k=1,4} 
Q_j^{}Q_k^{}\:v_C^{j,k}\:\tau_{z,i}^{}\:\bsigma_i^{}\cdot \kij{i}{}\\
&=&
{(1+\tau_{z,3}^{})\over 2}
\left[{(1+\tau_{z,1}^{})\over 2}\:v_C^{3,1}+{(1+\tau_{z,2}^{})\over 2}\:v_C^{3,2}\right] 
(\tau_{z,3}^{}\bsigma_3^{}\cdot\bfk_3^{}+\tau_{z,4}^{}\bsigma_4^{}\cdot\bfk_4^{})\non
&+& 
{(1+\tau_{z,4}^{})\over 2}
\left[{(1+\tau_{z,1}^{})\over 2}\:v_C^{4,1}+{(1+\tau_{z,2}^{})\over 2}\:v_C^{4,2}\right]
(\tau_{z,3}^{}\bsigma_3^{}\cdot\bfk_3^{}+\tau_{z,4}^{}\bsigma_4^{}\cdot\bfk_4^{})\non
&+&
{(1+\tau_{z,1}^{})\over 2}
\left[{(1+\tau_{z,3}^{})\over 2}\:v_C^{3,1}+{(1+\tau_{z,4}^{})\over 2}\:v_C^{4,1}\right]
(\tau_{z,1}^{}\bsigma_1^{}\cdot\bfk_1^{}+\tau_{z,2}^{}\bsigma_2^{}\cdot\bfk_2^{})\non
&+&
{(1+\tau_{z,2}^{})\over 2}
\left[{(1+\tau_{z,3}^{})\over 2}\:v_C^{3,2}+{(1+\tau_{z,4}^{})\over 2}\:v_C^{4,2}\right]
(\tau_{z,1}^{}\bsigma_1^{}\cdot\bfk_1^{}+\tau_{z,2}^{}\bsigma_2^{}\cdot\bfk_2^{})
\label{one1}.
\eea 
In the above equation, the one-body operator can either maintain the spin-parity-isospin
quantum numbers of a single deuteron, or produce a single two-nucleon state with $S=0$. As
the operators $v_C^{3,1},v_C^{3,2}$ do not flip the spin of their two-nucleon system, 
the quantum numbers of the $dd$ state must be maintained. Manipulations similar to those of
Sect.~\ref{eval} lead to the simplification
\bea
Y &=& \left[v_C^{3,1}+v_C^{3,2}-v_C^{4,1}-v_C^{4,2}\right]
\left(\frac{\bsigma_3^{}\cdot\bfk_3^{}-\bsigma_4^{}\cdot\bfk_4^{}}{4}\right)
\:+\: \left[v_C^{3,1}+v_C^{4,1}-v_C^{3,2}-v_C^{4,2}\right]
\left(\frac{\bsigma_1^{}\cdot\bfk_1^{}-\bsigma_2^{}\cdot\bfk_2^{}}{4}\right),
\eea
and finally to
\bea 
Y &\rightarrow& \left[v_C^{3,1}+v_C^{3,2}\right]\bfS_2^{}\cdot\bkappa_2^{}
\:+\: \left[v_C^{3,1}+v_C^{4,1}\right]\bfS_1^{}\cdot\bkappa_1^{},
\label{three1}
\eea   
which is analogous to that of Eq.~(\ref{three}) for the initial-state Coulomb interaction. 
Computation of the momentum space matrix element then leads to the result  
\bea
(\bfk,\bkappa_1,\bkappa_2\vert Y \vert dd,\bfp) &=& 
\frac{8\pi\alpha_{\mathrm{em}}^{}}{(2\pi)^{3/2} \bfv^2}\:
\left[\bfS_2\cdot\left(\bkappa_2+{\bfv\over2}\right)
-\bfS_1\cdot\left(\bkappa_1+{\bfv\over2}\right)\right]
\widetilde{\Phi}_d^{}\left(\bkappa_1^{}+\frac{\bfv}{2}\right)
\widetilde{\Phi}_d^{}\left(\bkappa_2^{}+\frac{\bfv}{2}\right).
\label{eight1}
\eea
Since $\bfp$ is again the only momentum remaining after integration, the above
matrix element may be treated along the same lines as Eq.~(\ref{eight}). Thus it is
again possible to extract a factor $(S_{2z}^{}-S_{1z}^{})$, giving finally
\bea
{\cal M}_C^F &=& \sqrt{6 E_\alpha^{} s}\:
\left(-\frac{\mu g_A^{}}{f_\pi^{}}\right)
\frac{8\pi\alpha_{\mathrm{em}}^{}}{(2\pi)^{3/2}}
\:\times\: M_2^{} \langle 1M_1^{},1M_2^{}\vert 0,0\rangle 
\:\times\: N_a^{}N_d^2
\:\times\: I_g^F\;\non
I_g^F &=& \int d^3kd^3\kappa_1^{}d^3\kappa_2^{}\:\:
\frac{(\bkappa_1^{}+\bkappa_2^{}+\bfv)\cdot\widehat{\bfp}}{\bfv^2}\:\: 
\left[-2M\epsilon_B^{}-(2\bkappa_1^2+2\bkappa_2^2+\bfk^2)\right]^{-1}_{}
\nonumber \\
&\times&
\exp{\left[-\frac{\alpha^2}{8}(\bfk^2+2\bkappa_1^2+2\bkappa_2^2)\right]} \:
\exp{\left[-\frac{\beta^2}{2}\left(\bkappa_1^{}+\frac{\bfv}{2}\right)^2\right]} \:
\exp{\left[-\frac{\beta^2}{2}\left(\bkappa_2^{}+\frac{\bfv}{2}\right)^2\right]}, 
\label{matel31}
\eea
where the Gaussian wave functions of Eqs.~(\ref{alpha}) and~(\ref{deut}) have
been employed. We proceed by writing the last two factors of \eq{matel3} in terms of Gaussians, and note that
the only difference with the general  procedure of the previous section is that we may use 
\bea
\left[-2M\epsilon_B^{}-(2\bkappa_1^2+2\bkappa_2^2+\bfk^2)\right]^{-1}_{} 
&=& -\int_0^\infty d\nu\:
\exp\left[-\nu\left(2M\epsilon_B^{}+2\bkappa_1^2+2\bkappa_2^2+\bfk^2\right)\right],
\label{extra1}
\eea
upon which $I^F_g$ becomes an 11-dimensional integral, of which 9~can again be
computed analytically by completion of squares in the exponents. Analogously to the
previous section, we employ the notation
\bea 
&&
\tilde\alpha^2_{}\equiv\alpha^2_{}+8\nu, \quad
\tilde R_l^2 \equiv \frac{\alpha^2}{2} + \beta^2 + 4\nu, \quad
\tilde R_v^2 \equiv \frac{\tilde R_l^2}{4} + \gamma - \frac{\beta^4}{4\tilde R_l^2}, \quad 
\eea 
along with the definition $I^F_j = |\bfp|\pi^{9/2}_{}I_{1j}^F$. The matrix element in
Eq.~(\ref{matel31}) then becomes
\bea
{\cal M}_C^F &=& \sqrt{6 E_\alpha^{} s}\:
\left(-\frac{\mu g_A^{}}{f_\pi^{}}\right)
\frac{8\pi\alpha_{\mathrm{em}}^{}}{(2\pi)^{3/2}}
\:\times\:
M_2^{} \langle 1M_1^{},1M_2^{}\vert 0,0\rangle 
\:\times\:
N_a^{}N_d^2\,
\:\times\:
|\bfp|\pi^{9/2}_{}
\:\times\:
I_{1g}^F, \non
I_{1g}^F &=& \int_0^\infty d\nu\,d\gamma\:
\frac{\tilde{\alpha}^4_{}}{2 \tilde R_l^8 \tilde R_v^5}\:
\exp\left[-\frac{\bfp^2\tilde\alpha^2_{}}{8}
\left(1-\frac{\tilde\alpha^2_{}}{8\tilde R_v^2}\right)\right]\:
\exp\left(-2\nu\,M\epsilon_B^{}\right).
\label{matel41}
\eea

The evaluation of the final-state Coulomb mechanism of the preceding subsection can also be implemented
using the Hulth\'en wave functions for the deuteron, given in~\eq{hulth}. The net result is that the integral
$I_{1g}^F$ should be replaced by $I_{1h}^F$, with
\bea 
I_{1h}^F &=&
\int_0^\infty d\nu\,d\gamma\,d\eta_1^{}d\eta_2^{}\:
\frac{\tilde{\alpha}^4_{}}{64 \tilde R_l^3 \tilde R_\kappa^5 \tilde R_v^5}\:
\exp\left[-\frac{\bfp^2\tilde\alpha^2_{}}{8}
\left(1-\frac{\tilde\alpha^2_{}}{8\tilde R_v^2}\right)\right]\:
\exp\left(-2\nu\,M\epsilon_B^{}\right)\:
f_{ab}^{}(\eta_1^{},\eta_2^{}),
\eea
where the definitions
\bea
&& \zeta_{12}^{}=\frac{\eta_1^{}+\eta_2^{}}{4}-\frac{(\eta_1^{}-\eta_2^{})^2}{4\tilde R_l^2},\quad 
\tilde R_l^2=\frac{\tilde{\alpha}^2}{2}+\eta_1^{}+\eta_2^{},\quad
\tilde R_\kappa^2=\frac{\tilde{\alpha}^2}{8}+\zeta_{12}^{},\quad
\tilde R_v^2=R_\kappa^2+\gamma-\frac{\zeta_{12}^2}{\tilde R_\kappa^2},
\eea
are used. The cross sections can then be computed using the expressions given in the previous
section on the initial-state Coulomb interaction.

\end{widetext}

\section{Numerical Results and Discussion}
\label{incorp}

The model parameters used in the present calculations are given in 
Table~\ref{tab_input}, and the calculated cross-sections at $T_d^{} = 228.5$~MeV 
and $T_d^{} = 231.8$~MeV are summarized in Table~\ref{tab_res}, such that 
$\sigma_g^{}$ and $\sigma_h^{}$ 
denote the results for the initial-state Coulomb interaction, obtained with 
Gaussian and Hulth\'en deuteron wave functions. If Gaussian wave 
functions are used throughout, the results are $59$~pb and $75$~pb at the two
energies considered. If Hulth\'en wave functions are used instead for the
deuterons, these results increase to $87$~pb and $111$~pb. In either case, 
the effects of the initial-state Coulomb interaction are significant, as the 
experimental values are $12.7$ and $15.1$~pb, respectively. It should also
be noted that these differences are much smaller than those encountered between
the \mbox{CD-Bonn} and Argonne~V18 potentials in Ref.~\cite{realistic}. The first 
toy-model calculations yielded nominal values of $23$~pb and $30$~pb. The present 
mechanism is therefore clearly large enough to warrant inclusion in a fully
realistic calculation.

\begin{table}[h]
\caption{Summary of parameters used in the calculation. The values of $\alpha$
and $\beta$, which appear in the expressions for the Gaussian bound-state wave functions,
are from Ref.~\cite{radii}, whereas $a$ and $b$ are relevant for the Hulth\'en deuteron,
and have been taken from Ref.~\cite{Tiburzi}.}
\vspace{.4cm}
\begin{tabular}{l||c}
$\alpha$~[fm] & $2.770$ \\
$\beta$~[fm] & $3.189$ \\
\vspace{-.3cm} & \\ \hline
\vspace{-.3cm} & \\
$a$~[fm$^{-1}_{}$] & $0.23161$ \\
$b$~[fm$^{-1}_{}$] & $1.3802$ \\
\vspace{-.3cm} & \\ \hline
\vspace{-.3cm} & \\
$M\epsilon_B^{}$[fm$^{-2}_{}$] & $0.68$ \\
$f_\pi^{}$~[MeV] & $92.4$ \\
$g_A^{}$ & $1.26$ \\
$\alpha_\mathrm{em}^{-1}$ & $137.04$ \\
\end{tabular}
\label{tab_input}
\end{table}

The results for the Coulomb interaction in the final state are denoted
$\sigma_g^F$ and $\sigma_h^F$, and are also given in Table~\ref{tab_res}. If Gaussian
wave functions are used throughout, the results are much smaller, about $1$~pb, which
represents $\sim 1$\% of those found for the Coulomb interaction in the initial state. 
The use of Hulth\'en deuteron wave functions in the initial state is found to enhance 
the effects of the final-state Coulomb interaction. However, they are still relatively 
small, about $10$\% of those of the initial-state Coulomb interaction. 

The principal result of this study is the manifest need to incorporate the effects of
Coulomb interactions in the initial state into the realistic calculation that includes
strong initial-state interactions. Recent progress in the treatment of Coulomb 
interactions in few-body scattering calculations~\cite{Deltuva:2006ch} should eventually 
allow such computations to be performed.

\begin{table}[t]
\caption{Summary of momenta, calculated cross-sections and integrals for the matrix elements. The quantities with
the subscripts $g$ and $h$ correspond to the Gaussian and Hulth\'en deuteron wave functions, respectively. The 
energies and momenta correspond to those of Ref.~\cite{IUCFCSB}.}
\vspace{.4cm}
\begin{tabular}{l||c|c}
& $T_d^{} = 228.5$~MeV & $T_d^{} = 231.8$~MeV \\
\vspace{-.3cm} && \\ \hline\hline
\vspace{-.3cm} && \\
$p$~[MeV] & $462.913$ & $466.924$ \\
$p_\pi^{}$~[MeV] & $19.372$ & $28.266$ \\
$E_\alpha^{}$~[MeV] & $3727.430$ & $3727.487$ \\
\vspace{-.3cm} && \\ \hline
\vspace{-.3cm} && \\
$I_{1g}^{}$~[fm$^{-5}_{}$] & $(1.3638, 1.3919)\times 10^{-5}_{}$ & $(1.2440, 1.3213)\times 10^{-5}_{}$ \\
$I_{1g}^F$~[fm$^{-5}_{}$] & $2.3043\times 10^{-6}_{}$ & $2.1713\times 10^{-6}_{}$ \\
$I_{1h}^{}$~[fm$^{-1}_{}$] & $(2.7011, 2.1617)\times 10^{-3}_{}$ & $(2.4919, 2.0821)\times 10^{-3}_{}$ \\
$I_{1h}^F$~[fm$^{-1}_{}$] & $1.1465\times 10^{-3}_{}$ & $1.0957\times 10^{-3}_{}$ \\
\vspace{-.3cm} && \\ \hline
\vspace{-.3cm} && \\
$\sigma_g^{}$~[pb] & $58.95$ & $75.25$ \\
$\sigma_h^{}$~[pb] & $85.77$ & $111.2$ \\
$\sigma_g^F$~[pb] & $0.824$ & $1.077$ \\
$\sigma_h^F$~[pb] & $9.419$ & $12.66$ \\
\end{tabular}
\label{tab_res}
\end{table}

\begin{acknowledgments}
We are grateful to Andrew Bacher, Edward Stephenson and Allena Opper 
for encouragement, many useful discussions, and for providing results from
the IUCF and TRIUMF experiments prior to publication. We also thank 
Christoph Hanhart and Antonio Fonseca for some valuable discussions.
This work was supported by the U.S. Department of Energy under grant
DE-FG-02-97ER41014. TL thanks the physics department of the University
of South Carolina for its hospitality during the completion of this work.
\end{acknowledgments}

\bibliographystyle{unsrt}

\end{document}